\documentclass[aps,prl,twocolumn,superscriptaddress,showpacs]{revtex4}
\usepackage{mathrsfs}
\usepackage{amssymb}
\usepackage{txfonts}
\usepackage{tipa}

\usepackage{graphicx}
\usepackage{dcolumn}
\usepackage{bm}

\begin{document}


\title{Field induced spin reorientation and giant spin-lattice coupling in EuFe$_2$As$_2$}

\author{Y. Xiao}
\email[y.xiao@fz-juelich.de]{}
\affiliation{Institut fuer Festkoerperforschung, Forschungszentrum
Juelich, D-52425 Juelich, Germany}

\author{Y. Su}
\affiliation{Juelich Centre for Neutron Science, IFF,
Forschungszentrum Juelich, Outstation at FRM II, Lichtenbergstr. 1,
D-85747 Garching, Germany}

\author{W. Schmidt}
\affiliation{Juelich Centre for Neutron Science, IFF,
Forschungszentrum Juelich, Outstation at Institut Laue-Langevin, BP
156, 38042 Grenoble Cedex 9, France}

\author{K. Schmalzl}
\affiliation{Juelich Centre for Neutron Science, IFF,
Forschungszentrum Juelich, Outstation at Institut Laue-Langevin, BP
156, 38042 Grenoble Cedex 9, France}

\author{C.M.N. Kumar}
\affiliation{Institut fuer Festkoerperforschung, Forschungszentrum
Juelich, D-52425 Juelich, Germany}

\author{S. Price}
\affiliation{Institut fuer Festkoerperforschung, Forschungszentrum
Juelich, D-52425 Juelich, Germany}

\author{T. Chatterji}
\affiliation{Juelich Centre for Neutron Science, IFF,
Forschungszentrum Juelich, Outstation at Institut Laue-Langevin, BP
156, 38042 Grenoble Cedex 9, France}

\author{R. Mittal}
\affiliation{Juelich Centre for Neutron Science, IFF,
Forschungszentrum Juelich, Outstation at FRM II, Lichtenbergstr. 1,
D-85747 Garching, Germany}

\affiliation{Solid State Physics Division, Bhabha Atomic Research
Centre, Trombay, Mumbai 400 085, India}

\author{L. J. Chang}
\affiliation{Nuclear Science and Technology Development Center,
National Tsing Hua University, Hsinchu 30013, Taiwan}

\author{S. Nandi}
\affiliation{Institut fuer Festkoerperforschung, Forschungszentrum
Juelich, D-52425 Juelich, Germany}

\author{N. Kumar}
\affiliation{Department of Condensed Matter Physics and Material
Sciences, Tata Institute of Fundamental Research, Homi Bhabha Road,
Colaba, Mumbai 400 005, India}

\author{S. K. Dhar}
\affiliation{Department of Condensed Matter Physics and Material
Sciences, Tata Institute of Fundamental Research, Homi Bhabha Road,
Colaba, Mumbai 400 005, India}

\author{A. Thamizhavel}
\affiliation{Department of Condensed Matter Physics and Material
Sciences, Tata Institute of Fundamental Research, Homi Bhabha Road,
Colaba, Mumbai 400 005, India}

\author{Th. Brueckel}
\affiliation{Institut fuer Festkoerperforschung, Forschungszentrum
Juelich, D-52425 Juelich, Germany} \affiliation{Juelich Centre for
Neutron Science, IFF, Forschungszentrum Juelich, Outstation at FRM
II, Lichtenbergstr. 1, D-85747 Garching, Germany}
\affiliation{Juelich Centre for Neutron Science, IFF,
Forschungszentrum Juelich, Outstation at Institut Laue-Langevin, BP
156, 38042 Grenoble Cedex 9, France}

\date{March 31, 2010}

\begin{abstract}

We have studied a EuFe$_2$As$_2$ single crystal by neutron
diffraction under magnetic fields up to 3.5 T and temperatures down
to 2 K. A field induced spin reorientation is observed in the
presence of a magnetic field along both the \emph{a} and \emph{c}
axes, respectively. Above critical field, the ground state
antiferromagnetic configuration of Eu$^{2+}$ moments transforms into
a ferromagnetic structure with moments along the applied field
direction. The magnetic phase diagram for Eu magnetic sublattice in
EuFe$_2$As$_2$ is presented. A considerable strain ($\sim$0.9\%) is
induced by the magnetic field, caused by the realignment of the
twinning structure. Furthermore, the realignment of the twinning
structure is found to be reversible with the rebound of magnetic
field, which suggested the existence of magnetic shape-memory
effect. The Eu moment ordering exhibits close relationship with the
twinning structure. We argue that the Zeeman energy in combined with
magnetic anisotropy energy is responsible for the observed
spin-lattice coupling.

\end{abstract}

\maketitle

The recent discovery of iron pnictide superconductors has triggered
extensive research on their physical properties and mechanism of
high temperature superconductors \cite{Kamihara,Chen1,Rotter1}. All
iron pnictides are found to be of layered structure in nature. For
undoped iron pnictides, the chains of parallel Fe spins within the
FeAs layers couple antiferromagnetically in the \emph{ab} plane of
the orthorhombic lattice with an antiparallel arrangement along the
\emph{c} axis \cite{Cruz,Huang,Su1}. This antiferromagnetic (AFM)
order in the parent compounds is likely due to a spin-density-wave
(SDW) instability caused by Fermi surface nesting \cite{Mazin}.
Similar to the high \emph{T}$_c$ cuprate superconductors, the
undoped iron pnictides are not superconducting under ambient
pressure and show an antiferromagnetic SDW order. Upon carrier
doping, the magnetic order is suppressed and superconductivity
emerges concomitantly \cite{Zhao1,Lester}.

EuFe$_2$As$_2$ is a peculiar member of the iron arsenide
\emph{A}Fe$_2$As$_2$ family since the \emph{A} site is occupied by
Eu$^{2+}$, which is an \emph{S}-state (orbital angular momentum
\emph{L} = 0) rare-earth ion possessing a 4\emph{f}$^7$ structure
with the total electron spin \emph{S} = 7/2. The theoretical
effective magnetic moment of Eu$^{2+}$ ion is 7.94 $\mu$$_B$. Our
previous neutron diffraction work on EuFe$_2$As$_2$ single crystals
revealed that the Fe and Eu spins form long range AFM order below
190 and 19 K, respectively \cite{Xiao1}. Furthermore, early studies
have shown that an external magnetic field may induce ferromagnetic
(FM) order in EuFe$_2$As$_2$ \cite{Jiang}, which suggested a weak
AFM coupling between Eu spins. The superconductivity can be achieved
by applying high pressure or doping on either Eu or As site in
EuFe$_2$As$_2$ \cite{Terashima,Ren1,Jeevan1,Qi}. Interestingly, some
works show that the ordering of the Eu$^{2+}$ moments persisted in
the superconducting phase, such as in the pressure-induced
EuFe$_2$As$_2$ superconductor \cite{Terashima} and
EuFe$_2$(As$_{0.7}$P$_{0.3}$)$_2$ superconductor \cite{Ren1}.
However, the role of Eu magnetism in superconductivity is not yet
clear and the Eu ordering state in different systems still need to
be clarified.

Here we report a single crystal neutron diffraction measurement on
EuFe$_2$As$_2$ under a magnetic field up to 3.5 T. The spin
reorientation of Eu moments is observed upon an applied magnetic
field parallel to both \emph{a} and \emph{c} axes of the
orthorhombic structure, while the Fe SDW order persists at high
magnetic fields. Interestingly, the application of a magnetic field
changes the twinning population in EuFe$_2$As$_2$ and the
redistribution of the domain population is found to be associated
with the evolution of the magnetic order of Eu moments, which
indicates the existence of a giant spin-lattice coupling effect. A
single crystal of EuFe$_2$As$_2$ was grown by the Sn-flux method
\cite{Xiao1}. It was in shape of a platelet with approximate
dimensions of 5~$\times$~5 $\times$~1~mm$^3$. Single crystal neutron
scattering measurements were performed on the thermal neutron
two-axis diffractometer D23 at the Institut Laue Langevin (Grenoble,
France). To investigate the evolution of the magnetic order of
EuFe$_2$As$_2$ under magnetic field, the crystal was aligned in the
\emph{a-c} scattering plane and a horizontal magnetic field up to
3.5 T was applied by using a cryomagnet. The measurement was
performed with an incident neutron wavelength of 1.28 $\buildrel
_\circ \over {\mathrm{A}}$.

\begin{figure}
\includegraphics[width=8.5cm,height=7cm]{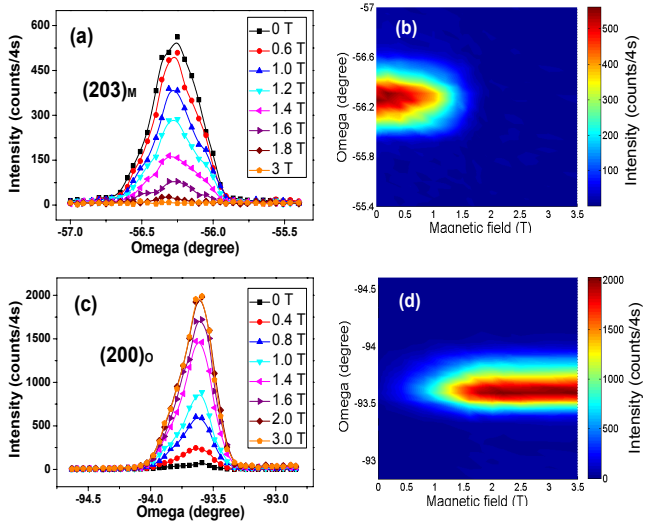}
\caption{\label{fig:epsart} (Color online) (a) Evolution of
(203)$_M$ magnetic reflections with the change of magnetic field at
2 K.(b) Magnetic field dependence of (203)$_M$ magnetic reflection
in two dimensional plot. (c) Evolution of (200)$_O$ reflection with
the change of magnetic field at 2 K. (d) Magnetic field dependence
of (200)$_O$ reflections in two dimensional plot.}
\end{figure}

In zero field, the crystal structure of EuFe$_2$As$_2$ can be well
described within the orthorhombic symmetry at 2 K and the magnetic
reflections originated from both Eu and Fe magnetic sublattices were
observed. The long range ordering of Eu$^{2+}$ [with \textbf{k} =
(0,0,1)] and Fe$^{2+}$ moments [with \textbf{k} = (1,0,1)] forms a
AFM structure as confirmed in our previous neutron diffraction study
\cite{Xiao1}. Several typical AFM reflections of the Eu ordering,
such as (201)$_M$, (203)$_M$ and (005)$_M$, were selected to examine
the magnetic structure evolution under the magnetic field (\emph{H})
along both \emph{c} and \emph{a} axes, i.e. [001] and [100]
directions, respectively. As an example, the field dependence of the
(203)$_M$ reflection is plotted in Fig. 1(a) and Fig. 1(b) in two
dimensions. Application of an field along \emph{c} axis strongly
weakens the (203)$_M$ reflection and suppresses it totally at the
critical field ($H_{Crit}^{Eu}$) of 1.8 T at 2 K. On the other hand,
the increase of intensity at some nuclear reflection positions, e.g.
(200)$_O$ as shown in Fig. 1(c) and (d), indicates that the Eu spin
gradually reorientate from the \emph{ab} plane to the [001]
direction. The field-induced magnetic phase transition takes place
from antiferromagnetic, via a canted configuration, to the
ferromagnetic structure. Note that the small nuclear structure
factor of the (200)$_O$ reflection makes it possible for us to
clearly observe the contribution from magnetic scattering.

\begin{figure}
\includegraphics[width=8.5cm,height=9cm]{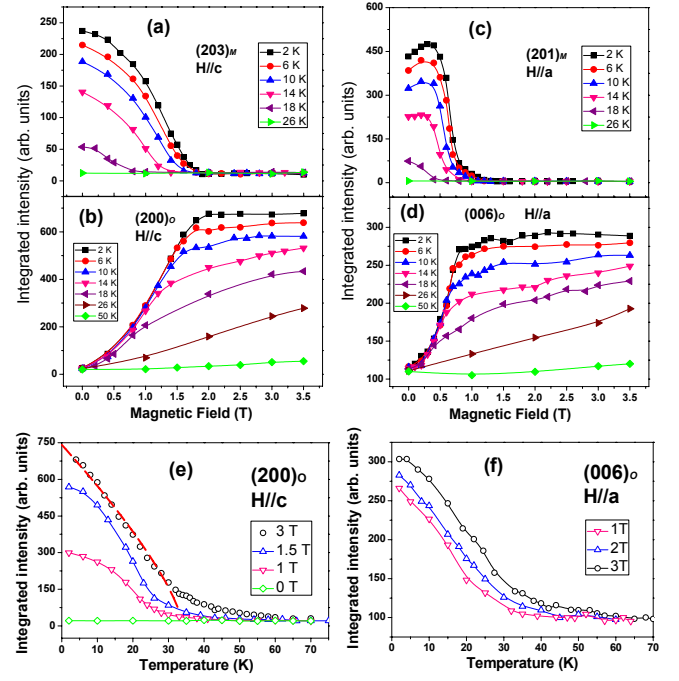}
\caption{\label{fig:epsart} (Color online) (a)(b) Integrated
intensity of (203)$_M$ and (200)$_O$ reflections as a function of
magnetic field with the field applied parallel to \emph{c} axis.
(c)(d) Magnetic field dependence of integrated intensity of
(201)$_M$ and (006)$_O$ reflections at different temperatures with
the magnetic field applied parallel to \emph{a} axis. (e)(f)
Temperature dependence of integrated intensity of (200)$_O$ and
(006)$_O$ reflections. }
\end{figure}

\begin{figure*}
\includegraphics[width=16cm,height=6.5cm]{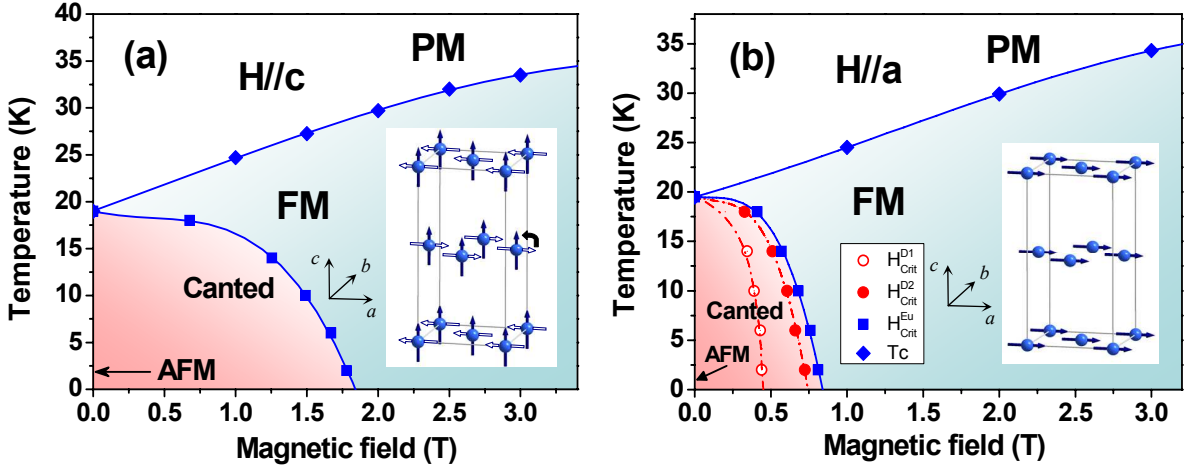}
\caption{\label{fig:wide}(Color online) (a) Magnetic phase diagram
for EuFe$_2$As$_2$ with applied field parallel to the
crystallographic \emph{c} axis. The inset shows the schematic view
of the magnetic structure. The AFM magnetic structure at zero field
is denoted as the open arrows, while the field induced FM structure
is denoted as the solid one. (b) Magnetic phase diagram for
EuFe$_2$As$_2$ with applied field parallel to the crystallographic
\emph{a} axis. The lines are guide to the eye. See text for more
details.}
\end{figure*}

\begin{figure}
\includegraphics[width=8.5cm,height=9.5cm]{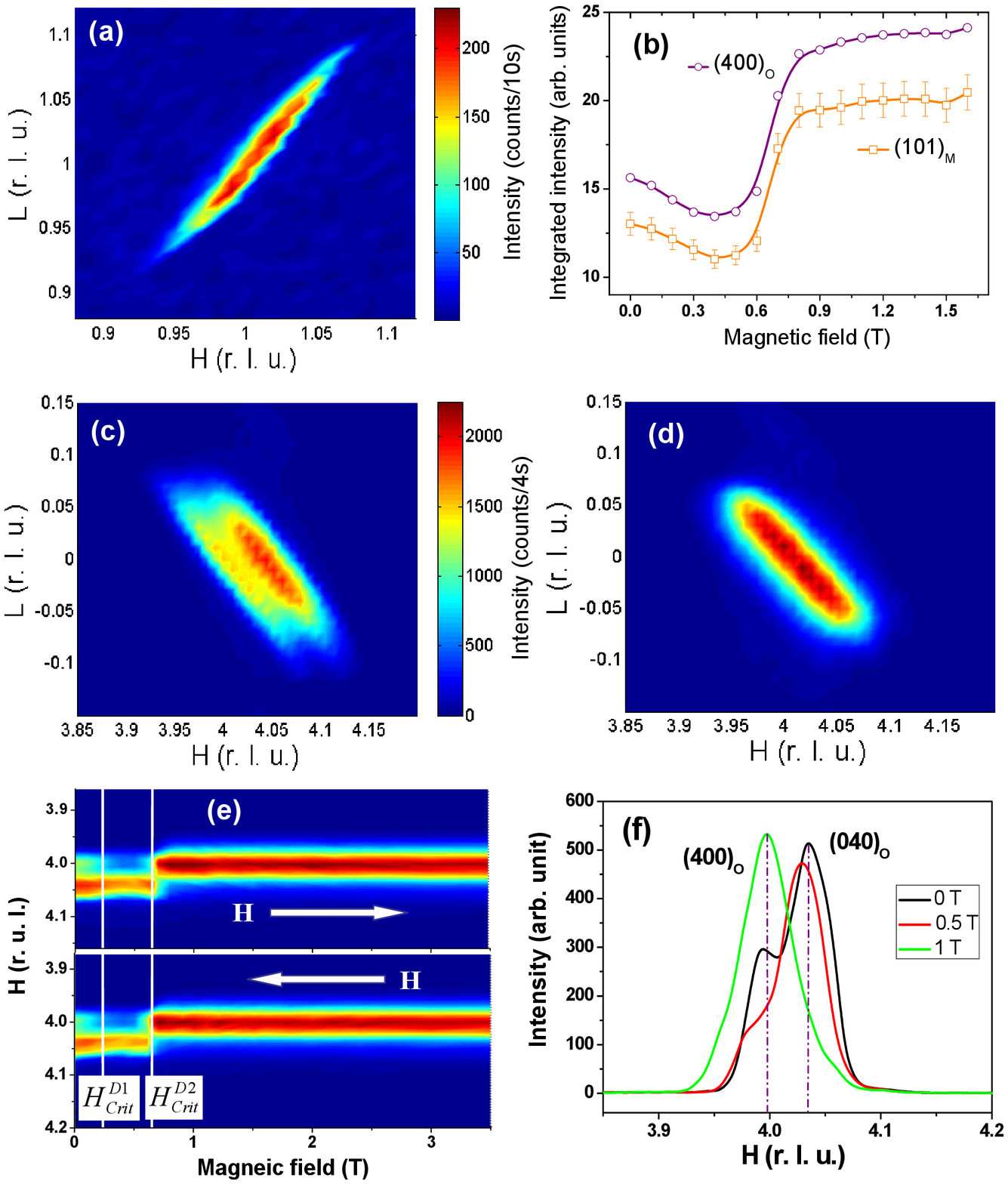}
\caption{\label{fig:epsart} (Color online) Selected Bragg
reflections of EuFe$_2$As$_2$ at 2 K. (a) The contour map of
(101)$_M$ reflection at 3 T field. (b) Field dependence of
integrated intensities of (400)$_O$ and (101)$_M$ reflections. (c)
and (d) The contour maps show the \emph{Q} dependence of the
(400)$_O$ and (040)$_O$ nuclear reflections at zero field and 3 T
field, respectively. (e) Evolution of the distribution of (400)$_O$
and (040)$_O$ reflections with increasing and decreasing magnetic
field at. (f) \emph{Q}-scans of (400)$_O$ reflection under three
typical magnitudes of magnetic field.}
\end{figure}

Fig. 2.(a) and (b) display the field dependence of integrated
intensities of selected magnetic scattering reflections in
EuFe$_2$As$_2$ with the field parallel to the \emph{c} axis at
different temperatures. The critical field at which the Eu magnetic
order changes from AFM to FM is decreasing with increasing
temperature. When the field is parallel to the \emph{a} axis, it is
observed that the intensity of the (201)$_M$ reflection increases
slightly and then decrease sharply with increasing strength of the
applied field at 2 K [see Fig. 2(c)]. It is known that diffraction
signals from both (201)$_M$ and (021)$_M$ reflections will be
detected due to the twinning configuration \cite{Xiao1}. The initial
increase of the (201)$_M$ intensity in Fig. 2(c) is caused by the
increasing contribution from the (021)$_M$ reflection, which is
related to the rotation of the (0\emph{k}0) domains. Actually, the
redistribution of domains under an applied field is observed and it
will be discussed in the following text. In Fig. 2(d), the increase
in the intensities of the (00\emph{l}) reflections with \emph{l}
even is observed instead of the (\emph{h}00) reflections. All these
results suggest that the Eu spins will orient to the direction of
the applied magnetic field once the field strength is greater than
the critical field. The FM arrangement of Eu spins is further
confirmed by the refinement of the reflections collected under the
field of 3 T at 2 K. The amplitude of Eu moment is estimated to be
6.9(4) $\mu$$_B$ and 6.7(4) $\mu$$_B$ for the two FM structures with
Eu moments along the \emph{c} and \emph{a} axis, respectively. The
temperature dependence of FM reflections was measured afterwards to
determine the ordering temperature of the Eu moments. Fig. 2(e) and
(f) show the temperature evolution of integrated intensities of the
(200)$_O$ and (006)$_O$ Bragg reflections for \emph{H} parallel to
\emph{c} and \emph{a} axes, respectively. A power-law was adopted to
describe this second order transition by fitting the temperature
variation in the ordering parameters. As indicated by the dash line,
the power-law fit on the data at 3 T in Fig. 2.(e) yields a magnetic
ordering transition temperature \emph{T}$_C$ = 33.5(4) K. The
intensity tail above \emph{T}$_C$ is caused by the contribution from
the field induced moment. According to the neutron measurement
results, the magnetic phase transition temperatures are determined
and plotted in Fig. 3(a) and (b). It can be seen that the
application of a magnetic field not only changes the ordering
configuration but also enhances the ordering temperature of the Eu
moment.

Besides of Eu order, we also examined the Fe order under the applied
field along both \emph{c} and \emph{a} axes. In contrast to the
reorientation of the Eu spins, the AFM SDW order of Fe is found to
be robust and it persists till fields up to 3 T. The two dimensional
plot of the (101)$_M$ reflection under a 3 T field at 2 K is shown
in Fig. 4(a). Surprisingly, it is found that the integrated
intensity of the (101)$_M$ magnetic reflection decreases slightly
and then sharply increases with increasing magnetic field, as shown
in Fig. 4(b). As we known that a twinning structure may exist in
orthorhombic EuFe$_2$As$_2$ phase due to the interchange of the
orthorhombic \emph{a} and \emph{b} axes. Additionally, the intensity
of the (101)$_M$ reflection only originates from the magnetic
scattering of the (\emph{h}00) twin. In order to clarify the
evolution of twinning structure under applied field, the (400)$_O$
reflection was examined from zero field up to 3.5 T since both
(040)$_O$ and (400)$_O$ reflections measured in present magnetic
structural configuration are pure nuclear reflections. As shown in
Fig. 4(c), the twinning structure forms at zero field and results in
the splitting between the (400)$_O$ and (040)$_O$ reflections. The
domain population is estimated to be 1:3 for the (\emph{h}00) and
(0\emph{k}0) twins at zero field. It is interesting that only the
(400)$_O$ reflection is observed when the magnetic field is
increased up to 3 T [Fig. 4(d)]. The single reflection from
(400)$_O$ is further confirmed by carefully checking the vicinity in
reciprocal space, which indicates that the single crystal is
detwinned. The detailed evolution of the domain population with
increasing field is illustrated by the intensity change of the
(400)$_O$ and (040)$_O$ reflections [Fig. 4(e)]. With increasing
field, the intensity of the (400)$_O$ decreases slowly and reaches
the lowest at the critical field $H_{Crit}^{D1}$ and suddenly
increases at the higher critical field $H_{Crit}^{D2}$. Whereas the
(040)$_O$ reflection vanishes at $H_{Crit}^{D2}$ in the other hand.
Therefore, $H_{Crit}^{D2}$ can be considered as the critical field
at which the (0\emph{k}0) domains overcome the domain wall energy
and realign to (\emph{h}00) domains. The realignment of the
(0\emph{k}0) domains will firstly introduce internal stress and this
stress will act on the (\emph{h}00) domains and tilt them slightly
away from the balance position, which leads to the decrease of the
(400)$_O$ reflection. However, (\emph{h}00) domains will rebound to
the balance position when the (0\emph{k}0) domains realign to the
(\emph{h}00) domains. Therefore, $H_{Crit}^{D1}$ corresponds to the
critical field when the (\emph{h}00) domains tilted and formed the
largest tilting angle away from the balance position. The
\emph{Q}-scans of the (400)$_O$ reflection under three typical
fields are plotted in Fig. 4(f). Based on the peak position of the
(400)$_O$ and (040)$_O$ reflections, the strain caused by the twin
boundary motion was estimated to be around 0.9\%. In contrast with
some magnetic shape-memory alloy, where the field induced strains
are associated with the martensite/austenite phase transition
\cite{Chmielus,Sozinov}, the field induced strains in the present
EuFe$_2$As$_2$ single crystal is caused by the domain realignment
purely. Furthermore, it is observed that the realignment of domains
is a reversible process with a rebound field, as shown in Fig.
4.(c).

The critical fields $H_{Crit}^{D1}$ and $H_{Crit}^{D2}$ are also
determined at different temperatures and plotted in Fig. 3 as the
red circles and dots. It is interesting that the critical field of
the domain redistribution is closely correlated with the field that
induces the AFM-FM transition of the Eu magnetic sublattice. The
energy difference between two domains seems strongly related with
the total energy of the Eu magnetism. Usually, the application of
mechanical compressive stress $\sigma$ on twinned single crystals
can induce a motion of the twin boundary, so called $'$detwinning$'$
process. For a twinned single crystal with FM structure, the
application of a magnetic field can also introduce stress which is
generated through the Zeeman energy \emph{E} =
-$\mu$$_0$$\vec{M}$$\cdot$$\vec{H}$ and magnetic anisotropy energy.
This detwinning process can be realized if the field induced stress
is larger than the internal stress originated from twin dislocation.
As illustrated in Fig. 3(b), the Eu spin rotates toward the \emph{a}
axis with increasing field. Concomitantly, the Zeeman energy
decreases with increasing field strength and magnetization of the
EuFe$_2$As$_2$ crystal. Thus, by minimizing the total energy,
greater stress is generated and results in a single domain. Since
the twinned structure only exist in the \emph{ab} plane in present
orthorhombic structure, the domain population does not change when a
magnetic field is applied parallel to the \emph{c} axis. Note that
the twinned structure also plays a key role with respect to the
magnetic and electric behaviors. For example, the in-plane
resistivity can be affected remarkably by twin boundary scattering
and the magnetization is associated with the response of different
domains to the applied field. Since the twinning structure exhibits
as the common feature for almost all orthorhombic phases in iron
pnictides \cite{Tanatar,Chu}, it is of great significant to detwin
the crystals before performing various experiments to evaluate the
real intrinsic properties.

Within the frame of the standard model for rare earth, the
crystalline electric field (CEF) effect is responsible for the
magnetic anisotropy for most of the rare earth ions with finite
orbital magnetic moment. However, for \emph{S}-state rare earth
ions, such as Gd$^{3+}$ and Eu$^{2+}$, the CEF effect is negligible
because of the vanishing of the orbital magnetic moment and the
charge density is no longer coupled to the spin. Nevertheless, weak
magnetic anisotropy is observed in both Gd$^{3+}$ and Eu$^{2+}$
contained compounds and the magnetic anisotropy is argued to be
driven mainly by dipole-dipole interactions
\cite{Colarieti,Laan,Abdelouahed}. The magnetic anisotropy energy
caused by the dipole-dipole interactions in EuFe$_2$As$_2$ was
evaluated for three different ordered states of Eu$^{2+}$ magnetic
moments: ground state AFM configuration, FM configurations with
moments aligned along \emph{a} and \emph{c} axes, respectively. By
considering the contribution of neighboring magnetic atoms within
the sphere of 25 $\buildrel _\circ \over {\mathrm{A}}$ radius, the
dipolar energies are obtained to be -206 $\mu$eV/Eu, -72 $\mu$eV/Eu
and 158 $\mu$eV/Eu for above mentioned three ordered configurations,
respectively. It clearly suggested that the dipolar interaction
favors the AFM ordered state, while the hard axis is predicted to be
\emph{c} axis. Our experimental data are in good agreement with the
prediction from the dipole-dipole interaction, which indicates a
dominant contribution of the dipole-dipole interactions to the
magnetic anisotropy of EuFe$_2$As$_2$. Given the fact that the
magnetization saturated at different fields for \emph{H} parallel to
the \emph{a} and \emph{c} axis due to the magnetic anisotropy
\cite{Jiang}, we argue that the magnetic anisotropy energy may also
generate stress and be partly responsible for the movement of the
twin boundaries under field.

In summary, our single crystal neutron diffraction experiments show
a magnetic field induced magnetic phase transition in EuFe$_2$As$_2$
with the Eu moments changing from AFM to FM arrangement. The
ordering temperature of the Eu moments increases with increasing
field. Moreover, giant spin-lattice coupling has been observed as
indicated by the redistribution of the domain population. Since the
domain realignment is intimately correlated with the magnetic phase
transition, the spin-lattice coupling in EuFe$_2$As$_2$ can be
attributed to the stresses generated by Zeeman energy and magnetic
anisotropy energy.

\appendix


\begin{thebibliography}{10}


\bibitem{Kamihara}
Y. Kamihara, T. Watanabe, M. Hirano, and H. Hosono, J. Am. Chem.
Soc. \textbf{130}, 3296 (2008).

\bibitem{Chen1}
X. H. Chen, T. Wu, G. Wu, R. H. Liu, H. Chen, and D. F. Fang, Nature
(London) \textbf{453}, 761 (2008).

\bibitem{Rotter1}
M. Rotter, M. Tegel, and D. Johrendt, Phys. Rev. Lett. \textbf{101},
107006 (2008).

\bibitem{Cruz}
C. de la Cruz, Q. Huang, J. W. Lynn, J. Li, W. Ratcliff II, J. L.
Zarestky, H. A. Mook, G. F. Chen, J. L. Luo, N. L. Wang, and P. C.
Dai, Nature (London) \textbf{453}, 899 (2008).


\bibitem{Huang}
Q. Huang, Y. Qiu, W. Bao, J. W. Lynn, M. A. Green, Y. Chen, T. Wu,
G. Wu, and X. H. Chen, Phys. Rev. Lett. \textbf{101}, 257003 (2008).

\bibitem{Su1}
Y. Su, P. Link, A. Schneidewind, Th. Wolf, Y. Xiao, R. Mittal, M.
Rotter, D. Johrendt, Th. Brueckel, and M. Loewenhaupt, Phys. Rev. B
\textbf{79}, 064504 (2009).

\bibitem{Mazin}
I. I. Mazin, D. J. Singh, M. D. Johannes, and M. H. Du, Phys. Rev.
Lett. \textbf{101}, 057003 (2008).


\bibitem{Zhao1}
J. Zhao, Q. Huang, C. de la Cruz, S. Li, J. W. Lynn, Y. Chen, M. A.
Green, G. F. Chen, G. Li, Z. Li, J. L. Luo, N. L. Wang, and P. Dai,
Nature Materials \textbf{7}, 953 (2008).


\bibitem{Lester}
C. Lester, Jiun-Haw Chu, J. G. Analytis, S. C. Capelli, A. S.
Erickson, C. L. Condron, M. F. Toney, I. R. Fisher, and S. M.
Hayden, Phys. Rev. B \textbf{79}, 144523 (2008).


\bibitem{Xiao1}
Y. Xiao, Y. Su, M. Meven, R. Mittal, C.M.N. Kumar, T. Chatterji, S.
Price, J. Persson, N. Kumar, S. K. Dhar, A. Thamizhavel, Th.
Brueckel, Phys. Rev. B \textbf{80}, 174424 (2009).


\bibitem{Jiang}
Shuai Jiang, Yongkang Luo, Zhi Ren, Zengwei Zhu, Cao Wang, Xiangfan
Xu, Qian Tao, Guanghan Cao and Zhu'an Xu, New Journal of Physics
\textbf{11}, 025007 (2008).



\bibitem{Terashima}
T. Terashima, M. Kimata, H. Satsukawa, A. Harada, K. Hazama, S. Uji, H. S. Suzuki, T. Matsumoto, and K. Murata, J. Phys. Soc. Jpn.
\textbf{78}, 083701 (2009).


\bibitem{Ren1}
Zhi Ren, Qian Tao, Shuai Jiang, Chunmu Feng, Cao Wang, Jianhui Dai,
Guanghan Cao, and Zhu'an Xu, Phys. Rev. Lett. \textbf{102}, 137002
(2009).

\bibitem{Jeevan1}
H. S. Jeevan, Z. Hossain, Deepa Kasinathan, Helge Rosner, C. Geibel,
and P. Gegenwart, Phys. Rev. B \textbf{78}, 092406 (2008).

\bibitem{Qi}
Yanpeng Qi, Zhaoshun Gao, Lei Wang, Dongliang Wang, Xianping Zhang,
and Yanwei Ma, New Journal of Physics \textbf{10}, 123003 (2008).


\bibitem{Chmielus}
M. Chmielus, X. X. Zhang, C. Witherspoon, D. C. Dunand and P.
M$\ddot{\texttt{u}}$llner, Nature Materials \textbf{8}, 863 (2009).


\bibitem{Sozinov}
A. Sozinov, A. A. Likhachev, N. Lanska and K. Ullakko, Appl. Phys.
Lett. \textbf{80}, 1746 (2002).



\bibitem{Tanatar}
M. A. Tanatar, E. C. Blomberg, A. Kreyssig, M. G. Kim, N. Ni, A.
Thaler, S. L. Bud'ko, P. C. Canfield, A. I. Goldman, I. I. Mazin,
and R. Prozorov, arXiv:1002.3801.

\bibitem{Chu}
Jiun-Haw Chu, James G. Analytis, Kristiaan De Greve, Peter L.
McMahon, Zahirul Islam, Yoshihisa Yamamoto, and Ian R. Fisher,
arXiv:1002.3364.


\bibitem{Colarieti}
M. Colarieti-Tosti, S. I. Simak, R. Ahuja, L.
Nordstr$\ddot{\texttt{o}}$m, O. Eriksson, D. $\buildrel _\circ
\over {\mathrm{A}}$berg, S. Edvardsson, and M. S. S. Brooks, Phys.
Rev. Lett. \textbf{91}, 157201 (2003).

\bibitem{Laan}
Gerrit van der Laan, Elke Arenholz, Andreas Schmehl, and Darrell G.
Schlom, Phys. Rev. Lett. \textbf{100}, 067403 (2008).

\bibitem{Abdelouahed}
Samir Abdelouahed and M. Alouani, Phys. Rev. B \textbf{79}, 054406
(2009).





\end{thebibliography}
\end{document}